# PART I : A SIMPLE SOLUTION OF THE TIME-INDEPENDENT SCHRÖDINGER EQUATION IN ONE DIMENSION

H. H. Erbil [a]

Ege University, Science Faculty, Physics Department     Bornova - IZMIR  35100, TURKEY

We found a simple procedure for the solution of the time-independent Schrödinger equation in one dimension without making any approximation. The wave functions are always periodic. Two difficulties may be encountered: one is to solve the equation $E = U(x)$, where E and U(x) are the total and potential energies, respectively, and the other is to calculate the integral $\int \sqrt{U(x)}\, dx$. If these calculations cannot be made analytically, it should then be performed by numerical methods. To find the energy and the wave function of the ground state, there is no need to calculate this integral, it is sufficient to find the classical turning points, that is to solve the equation $E = U(x)$.

## 1. INTRODUCTION

Although we succeed in solving the time-independent Schrödinger equation for some quantum mechanical problems in one dimension, an exact solution is not possible in complicated situations, and we must then resort to approximation methods. For the calculation of stationary states and energy eigenvalues, these include perturbation theory, the variation method and the WKB approximation. Perturbation theory is applicable if the Hamiltonian differs from an exactly solvable part by a small amount. The variation method is appropriate for the calculation of the ground state energy if one has a qualitative idea of the form of the wave function and the WKB method is applicable in the nearly classical limit. In this study we achieved a simple procedure for the exact solution of the time-independent Schrödinger equation in one dimension without making any approximation. We have applied this simple procedure to some quantum mechanical problems in one dimension in the following works.

## 2. TIME-INDEPENDENT SCHRÖDINGER EQUATION AND ITS SOLUTION IN ONE DIMENSION

The time-independent Schrödinger equation in one dimension is

$$\frac{d^2\psi(x)}{dx^2} + \frac{2m}{\hbar^2}[E - U(x)]\psi(x) = 0 \qquad (1)$$

Where, E and U(x) are the total (non relativistic) and potential energies of a particle of mass m, respectively. If $E > U(x)$, then the kinetic energy is positive (bound state and scattering) and If $E < U(x)$, then the kinetic energy is negative and not admissible classically (unbound state and tunnelling). These two cases is shown in Figure 1.

---

[a] E-mail: erbil@sci.ege.edu.tr     Fax: +90 232 388 1036

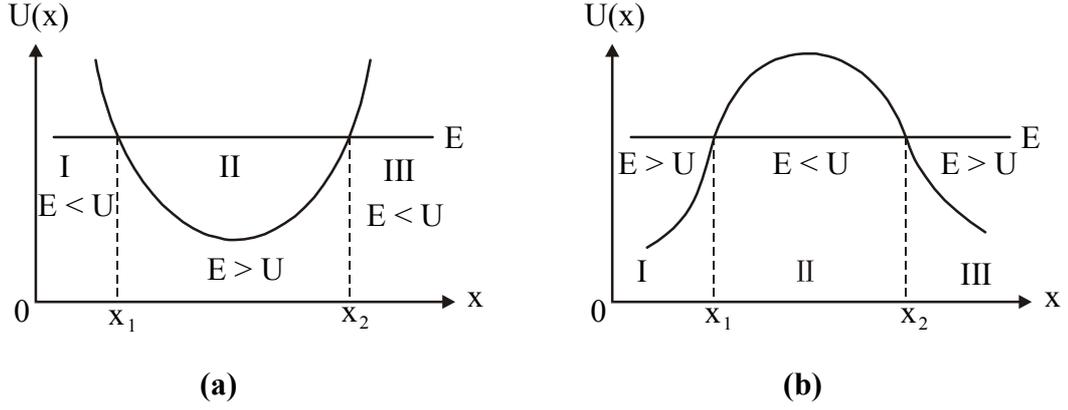

**Figure 1.** Domains relevant to a particle of energy E moving in a one dimensional potential field U(x): **(a)** In the domains I and III, E < U(x), thus, kinetic energies are negative (unbound state); In the domain II, E > U(x), thus, kinetic energy is positive (bound state). **(b)** In the domains I and III, E > U(x), thus, kinetic energies are positive: In the domain II, E < U(x), thus, kinetic energy is negative (unbound state). The roots of the equation E = U(x) are turning points of the corresponding classical motion.

## 2.1. The Case of E > U(x)   (kinetic energy is positive, bound state and scattering )

### 2. 1. 1. Solution for The Ground State

Now, we should solve the differential equation (1). To solve this equation for the ground state, let us perform the transformations below:

$$U(x) \rightarrow S\delta(x - x_0) \quad \text{and} \quad \psi(x) \rightarrow F(x) \tag{2}$$

Where $\quad S = \int_{x_1}^{x_2} U(x)\, dx \tag{3}$

Where $\delta(x-x_0)$ is Dirac function and $x_1$ and $x_2$ are the roots of the equation E = U(x). S is the area between the graph of U(x) and the x axis on [$x_1$, $x_2$]. If we take $x_0 = \dfrac{x_1 + x_2}{2}$ and $d = x_2 - x_1$, then we find $x_1 = x_0 - d/2$, $x_2 = x_0 + d/2$ (See Figure 1). With this transformation, the Schrödinger equation (1) becomes

$$\frac{d^2 F(x)}{dx^2} + \frac{2m}{\hbar^2} E F(x) = +\frac{2m}{\hbar^2} S\, \delta(x - x_0) F(x) \tag{4}$$

To evaluate the behaviour of F(x) at $x = x_0$, let us integrate the equation (4) over the interval ($x_0 - \varepsilon$, $x_0 + \varepsilon$) and let us consider the limit $\varepsilon \rightarrow 0$. We obtain

$$F'(x_0 + \varepsilon) - F'(x_0 - \varepsilon) = \frac{2m}{\hbar^2} S\, F(x_0) \tag{5}$$

This equation (5) shows that the derivation of F(x) is not continuous at the $x = x_0$ point [1,2]; whereas the wave function, F(x), should be continuous at the $x = x_0$ point.

To solve the differential equation (4), we can perform the transformation of Fourrier of the equation (4). FT [F(x)] = D(q) is Fourrier's Transformation of F(x). From (4):

$$FT\left[\frac{d^2F(x)}{dx^2}\right] + \frac{2m}{\hbar^2} E\, FT[F(x)] = +\frac{2m}{\hbar^2} S\, FT[F(x)\,\delta(x-x_0)]$$

$$-q^2 D(q) + \frac{2m}{\hbar^2} E\, D(q) = \frac{2m}{\hbar^2} S\, \frac{1}{\sqrt{2\pi}} \int_{-\infty}^{+\infty} e^{-iqx} F(x)\,\delta(x-x_0)\,dx$$

$$= \frac{2m}{\hbar^2} S\, \frac{1}{\sqrt{2\pi}} e^{-iqx_0} F(x_0)$$

From there, we get

$$D(q) = \frac{a^2\, e^{-iqx_0}}{q^2 + k_0^2} \tag{6}$$

Here, $k_0^2 = -\frac{2m}{\hbar^2} E$ and $a^2 = -\frac{2m}{\hbar^2} S\, \frac{1}{\sqrt{2\pi}} F(x_0)$

The function F(x) can be obtained by the inverse Fourrier's Transformation of D(q),

$$F(x) = FT^{-1}[D(q)] = \frac{a^2}{\sqrt{2\pi}} \int_{-\infty}^{+\infty} \frac{e^{-iqx_0}}{q^2 + k_0^2} e^{iqx}\, dq$$

$$= \frac{a^2}{\sqrt{2\pi}}\left\{\int_{-\infty}^{+\infty} \frac{\cos[q(x-x_0)]}{q^2+k_0^2}\,dq + i\int_{-\infty}^{+\infty} \frac{\sin[q(x-x_0)]}{q^2+k_0^2}\,dq\right\} = \frac{a^2}{|k_0|}\sqrt{\frac{\pi}{2}} e^{-|k_0||(x-x_0)|}$$

Or we can shortly write this function as follows:

$$F(x) = A e^{-k|(x-x_0)|} \tag{7}$$

Here, $k = |k_0|$ and $A = \frac{a^2}{k}\sqrt{\frac{\pi}{2}}$. From (7) we get

$$F(x) = A e^{k(x-x_0)} \quad \text{for} \quad x < x_0 \tag{8a}$$
$$F(x) = A e^{-k(x-x_0)} \quad \text{for} \quad x > x_0 \tag{8b}$$

Inserting the function (8) in the equation (5) and taking the limit $\varepsilon \to 0$, we obtain

$$k = \frac{m}{\hbar^2} S \quad \text{or} \quad E = -\frac{m}{2\hbar^2} S^2 \tag{9}$$

To find the constant A, the function F(x) can be normalized to 1:

$$\int_{-\infty}^{x_0} AA^* e^{2k(x-x_0)}\, dx + \int_{x_0}^{+\infty} AA^* e^{-2k(x-x_0)}\, dx = 1$$

From this equation we find

$$|A| = \sqrt{k} = \frac{\sqrt{mS}}{\hbar} \tag{10}$$

From (8a) and (8b), by adding and subtracting, we can get

$$F(x) = \frac{1}{2} A\left[e^{k(x-x_0)} + e^{-k(x-x_0)}\right] = A \cosh[k(x-x_0)] \tag{11a}$$

$$F(x) = \frac{1}{2}A\left[e^{k(x-x_0)} - e^{-k(x-x_0)}\right] = A\sinh[k(x-x_0)] \qquad (11b)$$

### 2.1.2. Solution For The Excited States

Now, to find the solution of the excited states, let us accept the wave function $\psi(x)$ to be $\psi(x) = F(x)\,e^{iG(x)}$ and we replace this function in the equation (1) and we get:

$$F''(x) - F(x)G'^{\,2}(x) - k^2 F(x) - m_1^2 U(x)F(x) + i\left[2F'(x)G'(x) + F(x)G''(x)\right] = 0 \qquad (12)$$

Here $m_1^2 = \dfrac{2m}{\hbar^2}$ and $k^2 = -\dfrac{2m}{\hbar^2}E$

The real and imaginary parts of the equation (12), we have two equations as follows:

$$F''(x) - F(x)G'^{\,2}(x) - k^2 F(x) - m_1^2 U(x)F(x) = 0 \qquad (13a)$$

$$2F'(x)G'(x) + F(x)G''(x) = 0 \qquad (13b)$$

To find the function $G(x)$, we use the equations (13). Let us take the function

$$F(x) = A\cosh[k(x-x_0)]$$

The first and second order derivatives of this function are as follows:

$$F'(x) = kA\sinh[k(x-x_0)] \quad \text{and} \quad F''(x) = k^2 A\cosh[k(x-x_0)] = k^2 F(x)$$

Inserting these functions into (13a), we get

$$k^2 F(x) - F(x)G'^{\,2}(x) - k^2 F(x) - m_1^2 U(x)F(x) = 0$$

Since $F(x) \neq 0$, dividing by $F(x)$, we have

$$G'^{\,2}(x) = -m_1^2 U(x) \quad \text{or} \qquad (14a)$$

$$G'(x) = \pm\sqrt{-m_1^2 U(x)} = \pm i\, m_1 \sqrt{U(x)} \qquad (14b)$$

From, the equation (14b), we get

$$G(x) = \pm i\, m_1 \int \sqrt{U(x)}\, dx + C \qquad (15)$$

Here C is any constant to be determined. On the other hand by derivation of the equation (14a), we can get $G''(x)$:

$$2G'(x)G''(x) = -m_1^2 U'(x) \quad \text{or} \quad G''(x) = -\frac{m_1^2}{2G'(x)}U'(x)$$

We replace this value of $G''(x)$ in (13b)

$$2F'(x)G'(x) - F(x)\frac{m_1^2}{2G'(x)}U'(x) = 0$$

Since $F'(x) = kA\sinh[k(x-x_0)]$, we get

$$[G'(x)]^2 = \frac{m_1^2}{4k}\coth[k(x-x_0)]U'(x) \qquad (16a)$$

$$G'(x) = \pm\frac{m_1}{2}\frac{1}{\sqrt{k}}\sqrt{U'(x)\coth[k(x-x_0)]} \qquad (16b)$$

Integrating the equation (16b), we have

$$G(x) = \pm \frac{m_1}{2} \frac{1}{\sqrt{k}} \int \sqrt{U'(x)\coth[k(x-x_0)]}\, dx + C \tag{17}$$

Multiplying the equation (13b) by F(x) we can rewrite it as follows:

$$\frac{d}{dx}[F^2(x)G'(x)] = 0 \quad \text{From this, we get} \quad F^2(x)G'(x) = \text{constant} = b \tag{18}$$

From the equation (18), we obtain

$$G'(x) = \frac{b}{F^2(x)} \tag{19}$$

Integrating the equation (19), we have

$$G(x) = \frac{b}{kA^2} \tanh[k(x-x_0)] + C \tag{20}$$

Finally, inserting $F'(x) = kA\sinh[k(x-x_0)]$ in the equation (13b) and integrating it we get

$$G(x) = \tanh[k(x-x_0)] + C \tag{21}$$

Comparing the equation (21) with the equation (20), we find that $b = kA^2$. Therefore, we have three functions for G(x). We rewrite these functions as follows:

$$G(x) = \pm i\, m_1 \int \sqrt{U(x)}\, dx + C \tag{22a}$$

$$G(x) = \pm \frac{m_1}{2} \frac{1}{\sqrt{k}} \int \sqrt{U'(x)\coth[k(x-x_0)]}\, dx + C \tag{22b}$$

$$G(x) = \tanh[k(x-x_0)] + C \tag{22c}$$

Thus, for the wave function $\psi(x)$, we have

$$\psi(x) = F(x)\left[Ae^{iG(x)} + Be^{-iG(x)}\right] \tag{23}$$

Here, A and B are constants to be determined by the boundary conditions.

Using the same procedure, for the other F(x) functions, we can obtain the different G(x) functions which are given in the Table I.

**Table I.** The elements of the wave functions in the case $E > U(x)$.

| F(x) | G(x) |
|---|---|
| $A\cosh[k(x-x_0)]$ | $\pm i\, m_1 \int \sqrt{U(x)}\, dx$ |
|  | $\pm \frac{m_1}{2} \frac{1}{\sqrt{k}} \int \sqrt{U'(x)\coth[k(x-x_0)]}\, dx$ |
|  | $\frac{b}{kA^2} \tanh[k(x-x_0)]$ |
|  | $\pm i\, m_1 \int \sqrt{U(x)}\, dx$ |
| $A \sinh[k(x-x_0)]$ | $\pm \dfrac{m_1}{2}\dfrac{1}{\sqrt{k}}\int \sqrt{U'(x)}\tanh[k(x-x_0)]\,dx$ |
| | $-\dfrac{b}{kA^2}\coth[k(x-x_0)]$ |
| $Ae^{k(x-x_0)}$ | $\pm i\,m_1 \int \sqrt{U(x)}\,dx$ |
| | $\pm \dfrac{m_1}{2}\dfrac{1}{\sqrt{k}}\int \sqrt{U'(x)}\,dx$ |
| | $-\dfrac{b}{2k}e^{-2k(x-x_0)}$ |
| $Ae^{-k(x-x_0)}$ | $\pm i\,m_1 \int \sqrt{U(x)}\,dx$ |
| | $\pm \dfrac{m_1}{2}\dfrac{i}{\sqrt{k}}\int \sqrt{U'(x)}\,dx$ |
| | $\dfrac{b}{2k}\dfrac{1}{F^2(x)}$ |
| $\psi(x)=F(x)\left[Ae^{iG(x)}+Be^{-iG(x)}\right]$ and $\psi(x)=Ae^{k(x-x_0)+iG(x)}+Be^{-k(x-x_0)-iG^*(x)}$ | |
| $F^2(x)G'(x)=\text{constant}=b$, $i=\sqrt{-1}$, $m_1=\sqrt{\dfrac{2m}{\hbar^2}}$, $k=\sqrt{-\dfrac{2m}{\hbar^2}E}$ | |
| If we take $kx$ instead of $k(x-x_0)$ in both $F(x)$ and $G(x)$, they are also solutions. | |

**2.2. The Case E < U(x)** (kinetic energy is negative and classically not admissible, unbound state)

To obtain the wave functions in this case, it is sufficient to replace $-k^2$ with $k^2$ (or $ik$ instead of $k$) and to use $-m_1^2$ instead of $m_1^2$ (or $i\,m_1$ instead of $m_1$) in the previous functions.

### 3. BOUNDARY CONDITIONS

Let us divide the potential domain into three parts, shown as in the Figure 1. In each domain, we show the functions $\psi_1(x)$, $\psi_2(x)$ and $\psi_3(x)$. The wave functions and their derivatives should be continuous. Because of these conditions the above functions must satisfy the following conditions.

$$\psi_1(x_1)=\psi_2(x_1) \qquad \psi'_1(x_1)=\psi'_2(x_1)$$
$$\psi_2(x_2)=\psi_3(x_2) \qquad \psi'_2(x_2)=\psi'_3(x_2)$$
(24)

The normalization of the bound state requires that the functions vanish at infinity. With these boundary conditions and normalization conditions of the wave functions, we can find the integral constants, A and B, and the energy E in the excited bound states. In the bound ground state, we do not need the solutions of the Schrödinger equation. It is sufficient to know only the classical turning points, $x_1$ and $x_2$. We will see it in the next section.

## 4. DETERMINATION OF THE GROUND STATE ENERGY IN THE BOUND STATE

The kinetic energy of the particle is

$$T = \frac{p^2}{2m} = E - U(x)$$

Integrating this equation from $x_1$ to $x_2$ and using the equation (3), we get

$$\int_{x_1}^{x_2} \frac{p^2}{2m} dx = \int_{x_1}^{x_2} [E - U(x)] dx = E(x_2 - x_1) - S$$

**(a)** For the positive kinetic energy, $E(x_2 - x_1) - S > 0$ (the bound state)
**(b)** For the negative kinetic energy, $E(x_2 - x_1) - S < 0$ (the unbound state).

As we can see the Figure 1, for the bound states, in the interval $[x_1, x_2]$, the kinetic energy is positive, outside the interval $[x_1, x_2]$, the kinetic energy is negative. The minimum point of the potential corresponds to ground state. Thus, at the minimum point of the potential, we can write.

$$E_0(x_2 - x_1) - S = 0$$

From this equation, we find the value of S as; $S = E_0(x_2 - x_1)$

Or, from Figure 1-a, we observe that we can also write $E(x_2 - x_1) - S = \int T \, dx$ for the region II, At ground state, the kinetic energy is zero, namely, T = 0. Therefore, from this equation we also find the same value $E_0(x_2 - x_1) = S$. We can replace this value of S in the equation (9) and we get

$$E = -\frac{m}{2\hbar^2} S^2 = -\frac{m}{2\hbar^2} E^2 (x_2 - x_1)^2 \quad \text{or} \quad E_0 = -\frac{2\hbar^2}{m} \frac{1}{(x_2 - x_1)^2} = -\frac{2\hbar^2}{m} \frac{1}{d^2} \quad (25)$$

$E_0$ represents ground state energy, where the negative sign indicates that the state is bound and it can be omitted it for the positive energies in the calculations.

Let us give an example to the ground state energy.

**Example**: $U(x) = \frac{1}{2} m\omega^2 x^2$ (the harmonic oscillator potential)

$$E = U(x) = \frac{1}{2} m\omega^2 x^2$$

From this equation $x_1 = -\sqrt{\frac{2E}{m\omega^2}}$ and $x_2 = \sqrt{\frac{2E}{m\omega^2}}$, $x_2 - x_1 = 2\sqrt{\frac{2E}{m\omega^2}}$

$$(x_2 - x_1)^2 = d^2 = 4\frac{2E}{m\omega^2}, \quad E_0 = \frac{2\hbar^2}{m} \frac{1}{(x_2 - x_1)^2} = \frac{2\hbar^2}{m} \frac{m\omega^2}{2.4E_0} = \frac{\hbar^2 \omega^2}{4E_0}$$

$$E_0^2 = \frac{1}{4} \hbar^2 \omega^2 \quad \rightarrow \quad E_0 = \frac{1}{2} \hbar\omega \quad \text{which is the ground state energy of the harmonic oscillator.}$$

## 5. CONCLUSION

Examining the Table I. in detail, we can receive the main solutions of Schrödinger equation in one dimension as follows:

1) $\psi(x) = e^{k(x-x_0)} \left[ A e^{iG(x)} + B e^{-iG(x)} \right]$

2) $\psi(x) = e^{-k(x-x_0)}\left[Ae^{iG(x)} + Be^{-iG(x)}\right]$

3) $\psi(x) = \cosh[k(x-x_0)]\left[Ae^{iG(x)} + Be^{-iG(x)}\right]$

4) $\psi(x) = \sinh[k(x-x_0)]\left[Ae^{iG(x)} + Be^{-iG(x)}\right]$

5) $\psi(x) = Ae^{k(x-x_0)+iG(x)} + Be^{-k(x-x_0)-iG^*(x)}$

If one takes kx instead of $k(x-x_0)$ in the above functions, they are also solutions.

Where, (a) For $E > U(x)$, $k = im_1\sqrt{E}$, $G(x) = im_1\int\sqrt{U(x)}dx$

(b) For $E < U(x)$, $k = m_1\sqrt{E}$, $G(x) = m_1\int\sqrt{U(x)}dx$

Here, $m_1 = \sqrt{\dfrac{2m}{\hbar^2}}$, $x_1$ and $x_2$ are the roots of the equation $E = U(x)$, $x_0 = \dfrac{x_1+x_2}{2}$, $d = x_2 - x_1$, $i = \sqrt{-1}$, $x_1 = x_0 - d/2$, $x_2 = x_0 + d/2$

The bound ground state energy is given by the formula

$$E_0 = -\frac{2\hbar^2}{m}\frac{1}{(x_2-x_1)^2} = -\frac{2\hbar^2}{m}\frac{1}{d^2}.$$

The negative sign of this formula shows the bound state and it can be omitted it for the positive energies in the calculations. For the bound ground state wave function, it should be taken $G(x) = 0$ and $k = k_0 = m_1\sqrt{E_0}$ in the above functions.

The functions 3) and 4) include the functions 1) and 2), because 3) and 4) are the linear combinations of 1) and 2). The functions 5) include also the functions 3) and 4). Thus, it is possible to take only the function 5) as the solution of Schrödinger equation in one dimension. A and B coefficients are determined by using the boundary and normalization conditions. Although the function 5) is sufficient by itself, we nevertheless use the other functions ( 1) to 5) ) in some applications.

In our procedure, there are two difficulties: one is to solve the equation $E = U(x)$ and the other one is to integrate $\sqrt{U(x)}$, namely, to do the calculation of $\int\sqrt{U(x)}\,dx$. If one can not do these calculations analytically, they must be performed by numerical methods. To find the energy and the wave function of the ground state, there is no need to calculate this integral, it is sufficient to find the classical turning points, namely, to solve the equation $E = U(x)$.

### Acknowledgements

I would like to express my sincere gratitude to my wife Özel and my daughters Işıl and Beril for their help and patience during my study as well as my colleague Prof.N.Armağan for his help.

# PART II : SOLUTION OF THE SCHRÖDINGER EQUATION IN ONE DIMENSION BY A SIMPLE PROCEDURE FOR A PARTICLE IN AN INFINITELY HIGH POTENTIAL WELL OF ARBITRARY FORM AND SOME EXAMPLES

H. H. Erbil [a]

Ege University, Science Faculty, Physics Department    Bornova - IZMIR  35100, TURKEY

The Schrödinger Equation for a particle in an infinitely high potential well of arbitrary form was solved by employing a simple procedure without making any approximation. Two different solutions were obtained, namely, symmetric and antisymmetric. These solutions are always periodic functions.

Keywords :  Infinitely high potential wells

## 1. INTRODUCTION

The time-independent Schrödinger equation in one dimension is given as follows:

$$\frac{d^2\psi(x)}{dx^2} + \frac{2m}{\hbar^2}[E - U(x)]\psi(x) = 0$$

where, E and U(x) are the total and potential energies, respectively. In the previous study [1], we have generally solved these equations and we have obtained the following functions:

1) $\psi(x) = e^{k(x-x_0)}\left[Ae^{iG(x)} + Be^{-iG(x)}\right]$

2) $\psi(x) = e^{-k(x-x_0)}\left[Ae^{iG(x)} + Be^{-iG(x)}\right]$

3) $\psi(x) = \cosh[k(x-x_0)]\left[Ae^{iG(x)} + Be^{-iG(x)}\right]$

4) $\psi(x) = \sinh[k(x-x_0)]\left[Ae^{iG(x)} + Be^{-iG(x)}\right]$

5) $\psi(x) = Ae^{k(x-x_0)+iG(x)} + Be^{-k(x-x_0)-iG^*(x)}$

Where,     (a) For $E > U(x)$, $k = i m_1 \sqrt{E}$,     $G(x) = i m_1 \int \sqrt{U(x)}\, dx$

(b) For $E < U(x)$, $k = m_1 \sqrt{E}$,     $G(x) = m_1 \int \sqrt{U(x)}\, dx$

$x_1$ and $x_2$ are the roots of the equation $E = U(x)$, $x_0 = \frac{x_1 + x_2}{2}$, $d = x_2 - x_1$, $x_1 = x_0 - d/2$, $x_2 = x_0 + d/2$, $m_1 = \sqrt{\frac{2m}{\hbar^2}}$. The bound ground state energy is given by the formula

---

[a] E-mail: erbil@sci.ege.edu.tr    Fax : +90 232 388 1036

$$E_0 = -\frac{2\hbar^2}{m}\frac{1}{(x_2-x_1)^2} = -\frac{2\hbar^2}{m}\frac{1}{d^2}$$

The negative sign appearing at the right-hand side in this formula shows the bound state and it can be omitt[ed] functions, they are also solutions. A and B are constant to be determined by the boundaries and normalization conditions.

## 2. SOLUTION OF THE SCHRÖDINGER EQUATION

### 2.1. WAVE FUNCTIONS

We consider the potential energy of the particle, U(x), so that $U(x) \geq 0$, $\lim_{x\to\pm\infty} U(x) = +\infty$. This potential is depicted in Figure 1.

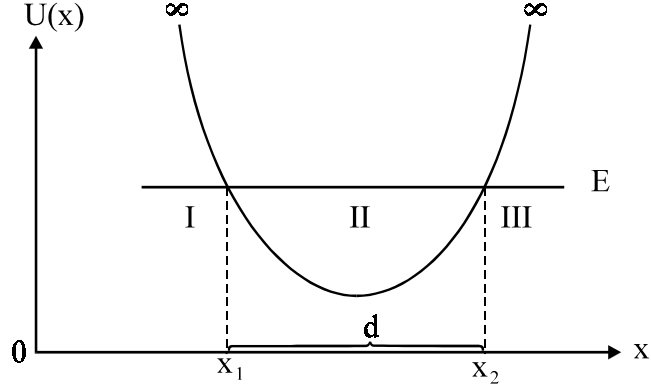

**Figure 1.** An infinitely high potential well; $x_1$ and $x_2$ are the roots of $U(x) = E$, $d = x_2-x_1$.

According to Figure 1, in the domains I and III, the potentials are infinite. Therefore, the corresponding wave functions must vanish in these domains: namely, $\psi_1(x) = 0$ and $\psi_3(x) = 0$. In the domain II, the wave function is different from zero, namely, $\psi_2(x) = \psi(x) \neq 0$. In this domain, $E > U(x)$ and $E > 0$. Thus,

$$G(x) = i\,m_1 \int \sqrt{U(x)}\,dx = i\,Q(x), \text{ with } Q(x) = m_1 \int \sqrt{U(x)}\,dx$$

We suppose that Q(x) is real, then $G^*(x) = -iQ(x)$, $k = i\,m_1\sqrt{E} = iK$, with $K = m_1\sqrt{E}$

1) $\psi(x) = e^{k(x-x_0)}\left[A_1 e^{iG(x)} + B_1 e^{-iG(x)}\right]$

$= e^{iK(x-x_0)}\left[A_1 e^{i\cdot iQ(x)} + B_1 e^{-i\cdot iQ(x)}\right] = e^{iK(x-x_0)}\left[A_1 e^{-Q(x)} + B_1 e^{+Q(x)}\right]$.

In order that $\psi(x)$ must vanish for $x = \pm\infty$; if $Q(x) > 0$, $B_1$ must be identically zero and if $Q(x) < 0$, $A_1$ must be identically zero. We suppose that $Q(x) > 0$. Therefore,

$$\psi(x) = A_1 e^{iK(x-x_0)}e^{-Q(x)} = \{A\cos[K(x-x_0)] + B\sin[K(x-x_0)]\}e^{-Q(x)} \quad (1a)$$

2) $\psi(x) = e^{-k(x-x_0)}\left[A_1 e^{iG(x)} + B_1 e^{-iG(x)}\right]$

$$= e^{-iK(x-x_0)}\left[A_1 e^{i\cdot iQ(x)} + B_1 e^{-i\cdot iQ(x)}\right] = e^{-iK(x-x_0)}\left[A_1 e^{-Q(x)} + B_1 e^{Q(x)}\right].$$

In order that $\psi(x)$ must vanish for $x = \pm\infty$; if $Q(x) > 0$, $B_1$ must be identically zero and if $Q(x) < 0$, $A_1$ must be identically zero. We suppose that $Q(x) > 0$. Therefore,

$$\psi(x) = A_1 e^{-iK(x-x_0)} e^{-Q(x)} = \{A\cos[K(x-x_0)] + B\sin[K(x-x_0)]\} e^{-Q(x)} \tag{1b}$$

3) $\psi(x) = \cosh[k(x-x_0)]\left[A_1 e^{iG(x)} + B_1 e^{-iG(x)}\right]$

$$= \cosh[iK(x-x_0)]\left[A_1 e^{i\cdot iQ(x)} + B_1 e^{-i\cdot iQ(x)}\right] = \cos[K(x-x_0)]\left[A_1 e^{-Q(x)} + B_1 e^{Q(x)}\right].$$

In order that $\psi(x)$ must vanish for $x = \pm\infty$; if $Q(x) > 0$, $B_1$ must be identically zero and if $Q(x) < 0$, $A_1$ must be identically zero. We suppose that $Q(x) > 0$. Therefore,

$$\psi(x) = A\cos[K(x-x_0)]e^{-Q(x)} \tag{1c}$$

4) $\psi(x) = \sinh[k(x-x_0)]\left[A_1 e^{iG(x)} + B_1 e^{-iG(x)}\right]$

$$= \sinh[iK(x-x_0)]\left[A_1 e^{i\cdot iQ(x)} + B_1 e^{-i\cdot iQ(x)}\right] = i\sin[K(x-x_0)]\left[A_1 e^{-Q(x)} + B_1 e^{Q(x)}\right].$$

In order that $\psi(x)$ must vanish for $x = \pm\infty$; if $Q(x) > 0$, $B_1$ must be identically zero and if $Q(x) < 0$, $A_1$ must be identically zero. We suppose that $Q(x) > 0$. Therefore, with $B = iA_1$

$$\psi(x) = B\sin[K(x-x_0)]e^{-Q(x)} \tag{1d}$$

5) $\psi(x) = A_1 e^{k(x-x_0)+iG(x)} + B_1 e^{-k(x-x_0)-iG^*(x)}$

$$= A_1 e^{iK(x-x_0)+i\cdot iQ(x)} + B_1 e^{-iK(x-x_0)-i\cdot(-i)Q(x)} = \left[A_1 e^{iK(x-x_0)} + B_1 e^{-iK(x-x_0)}\right] e^{-Q(x)}$$

$$= \{A\cos[K(x-x_0)] + B\sin[K(x-x_0)]\} e^{-Q(x)} \tag{1e}$$

The functions (1a), (1b) and (1e) are the same. They are also linear combinations of (1c) and (1d). Thus, we can consider one of them and apply the boundary conditions. Boundary conditions are,

$$\psi(x_1) = 0, \quad \psi(x_2) = 0$$

$$\psi(x) = \{A\cos[K(x-x_0)] + B\sin[K(x-x_0)]\} e^{-Q(x)}$$

$$\psi(x_1) = \{A\cos[K(x_1-x_0)] + B\sin[K(x_1-x_0)]\} e^{-Q(x_1)} = 0$$

$$\psi(x_2) = \{A\cos[K(x_2-x_0)] + B\sin[K(x_2-x_0)]\} e^{-Q(x_2)} = 0$$

$$x_1 = x_0 - d/2, \quad x_2 = x_0 + d/2, \quad e^{-Q(x_1)} \neq 0, \quad e^{-Q(x_2)} \neq 0$$

So, $\quad A\cos(Kd/2) - B\sin(Kd/2) = 0 \tag{2a}$

$\quad A\cos(Kd/2) + B\sin(Kd/2) = 0 \tag{2b}$

In order that this system of equations has a solution different from zero, the determinant of coefficients should vanish, namely,

$$\det\begin{vmatrix} \cos(Kd/2) & -\sin(Kd/2) \\ \cos(Kd/2) & \sin(Kd/2) \end{vmatrix} = 0 \quad \text{or} \quad \cos(Kd/2)\cdot\sin(Kd/2) = 0 \tag{3}$$

From the equation (3) we have:

**(a)** $\cos(Kd/2) = 0$ and $\sin(Kd/2) \neq 0$

$$\frac{Kd}{2} = (2n-1)\frac{\pi}{2} \quad \text{or} \quad Kd = (2n-1)\pi, \quad n = 1, 2, 3, ... \quad (4a)$$

Replacing this value of Kd in (2), we obtain $B \equiv 0$. Thus, the function should be

$$\psi_n(x) = A\cos[K(x-x_0)]e^{-Q(x)} = A\cos\left[\frac{(2n-1)\pi}{d}(x-x_0)\right]e^{-Q(x)}, \quad n = 1, 2, 3, ... \quad (4b)$$

This is a symmetric function.

**(b)** $\cos(Kd/2) \neq 0$ and $\sin(Kd/2) = 0$

$$\frac{Kd}{2} = n\pi \quad \text{or} \quad Kd = 2\pi n, \quad n = 1, 2, 3, ... \quad (5a)$$

Inserting this value of Kd into (2), we obtain $A \equiv 0$. Thus, the function should be

$$\psi_n(x) = B\sin[K(x-x_0)]e^{-Q(x)} = B\sin\left[\frac{2\pi n}{d}(x-x_0)\right]e^{-Q(x)}, \quad n = 1, 2, 3, ... \quad (5b)$$

This is an antisymmetric function.

**(c)** From the equation (3) we obtain

$$\sin(Kd) = 0 \quad \text{or} \quad Kd = n\pi, \quad n = 1, 2, 3, ... \quad (6a)$$

Substituting this value of Kd into (2), we obtain $A = 0$, $B \neq 0$ for even integer n values; and $A \neq 0$, $B = 0$ for odd integer n values. Thus, we have two wave functions as follows:

$$\psi_n(x) = A\cos\left[\frac{n\pi}{d}(x-x_0)\right]e^{-Q(x)} \quad \text{for odd integer n values} \quad (6b)$$

$$\psi_n(x) = B\sin\left[\frac{n\pi}{d}(x-x_0)\right]e^{-Q(x)} \quad \text{for even integer n values} \quad (6c)$$

Since odd and even integers can be written as (2n-1) and (2n), respectively, we again obtain the wave functions as

$$\psi_n(x) = A\cos\left[\frac{(2n-1)\pi}{d}(x-x_0)\right]e^{-Q(x)}, \quad n = 1, 2, 3, ... \quad (4b')$$

$$\psi_n(x) = B\sin\left[\frac{2n\pi}{d}(x-x_0)\right]e^{-Q(x)}, \quad n = 1, 2, 3, ... \quad (5b')$$

The function (4b) is the same as (1c) and the function (5b) is the same as (1d). Therefore, we have in fact two functions. One of them is symmetric, the other is antisymmetric and they can be given by (4b) and (5b), respectively. We can find the coefficients A and B by the normalisation of the wave functions. Thus, the wave functions can be obtained as

$$\psi_n(x) = A\cos\left[\frac{(2n-1)\pi}{d}(x-x_0)\right]e^{-Q(x)}, \quad n = 1, 2, 3, ... \text{ (symmetric function)} \quad (7a)$$

$$\psi_n(x) = B\sin\left[\frac{2\pi n}{d}(x-x_0)\right]e^{-Q(x)}, \quad n = 1, 2, 3, ... \text{ (antisymmetric function)} \quad (7b)$$

## 2.2. ENERGY VALUES

### 2.2.1. Ground State Energy

To determine the ground state energy (or zero point energy) we can use the formula

$$E_0 = \frac{2\hbar^2}{m} \frac{1}{(x_2 - x_1)^2} = \frac{2\hbar^2}{m} \frac{1}{d^2} \tag{8}$$

### 2.2.2. Excited State Energies

To determine the excited state energies, we can use the equations (4a), (5a) and (6a)

**(a)** From (4a), we can have the symmetric state energies

$$K^2 d^2 = (2n-1)^2 \pi^2 = 4\pi^2 \left(n - \tfrac{1}{2}\right)^2, \qquad \frac{2m}{\hbar^2} E_n \cdot d^2 = 4\pi^2 \left(n - \tfrac{1}{2}\right)^2$$

$$E_n = \frac{2\hbar^2 \pi^2}{md^2} \left(n - \tfrac{1}{2}\right)^2, \qquad n = 1, 2, 3, \ldots \tag{9a}$$

**(b)** From (5a), we can have the antisymmetric state energies

$$K^2 d^2 = 4\pi^2 n^2, \qquad \frac{2m}{\hbar^2} E_n d^2 = 4\pi^2 n^2, \qquad E_n = \frac{2\hbar^2 \pi^2}{md^2} n^2, \quad n = 1, 2, 3, \ldots \tag{9b}$$

**(c)** From (6a), we can have the general case energies

$$K^2 d^2 = \pi^2 n^2, \qquad \frac{2m}{\hbar^2} E_n d^2 = \pi^2 n^2, \qquad E_n = \frac{\hbar^2 \pi^2}{2md^2} n^2, \quad n = 1, 2, 3, \ldots \tag{9c}$$

From (9c), for odd n values, we obtain (9a) and for even n values, we get (9b). Thus, the energy values (9c) include both (9a) and (9b).

## 3. EXAMPLES

### Example 1. Infinitely High Square Potential Well

We consider a particle of mass m captured in a box limited by $0 \leq x \leq L$. The corresponding potential is given by

$U(x) = 0$     for     $0 < x < L$

$U(x) = \infty$     for     $x < 0$ and $x > L$     (see Figure 2)

Outside the potential well the wave functions must vanish, because $U(x)$ is infinitely large. Thus, in the domain I and III, the wave functions are zero, namely, $\psi_1(x) = 0$ and $\psi_3(x) = 0$. In the domain II, the wave function $\psi_2(x) = \psi(x) \neq 0$.

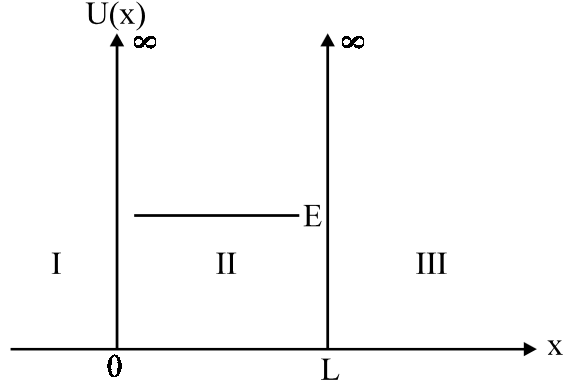

**Figure 2.** The one dimensional infinitely high square potential well.

In the domain II, $U(x) = 0$, $E > 0$, $x_1 = 0$, $x_2 = L$, $x_0 = \dfrac{x_1 + x_2}{2} = \dfrac{L}{2}$, $d = x_2 - x_1 = L$, $x_1 = x_0 - d/2$, $x_2 = x_0 + d/2$, $G(x) = 0$ and $Q(x) = 0$ because $U(x) = 0$.

### Ex-1.1. Wave Functions

From the Equations (7a) and (7b), we obtain:

$$\psi_n(x) = A\cos\left[\frac{(2n-1)\pi}{L}(x - \frac{L}{2})\right] \quad , \quad n = 1, 2, 3, \ldots \text{ (symmetric function)}$$

$$\psi_n(x) = B\sin\left[\frac{2n\pi}{L}(x - \frac{L}{2})\right] \quad , \quad n = 1, 2, 3, \ldots \text{ (antisymmetric function)}$$

We can find the coefficients A and B by the normalisation of the wave functions. After an easy calculation, we get

$$|A| = \sqrt{\frac{2}{L}} \quad \text{and} \quad |B| = \sqrt{\frac{2}{L}}$$

Thus, the wave functions can be obtained as

$$\psi_n(x) = \sqrt{\frac{2}{L}}\cos\left[\frac{(2n-1)\pi}{L}(x - \frac{L}{2})\right] \quad , \quad n = 1, 2, 3, \ldots \text{(symmetric function)} \quad \text{(E1-1a)}$$

$$\psi_n(x) = \sqrt{\frac{2}{L}}\sin\left[\frac{2n\pi}{L}(x - \frac{L}{2})\right] \quad , \quad n = 1, 2, 3, \ldots \text{(antisymmetric function)} \quad \text{(E1-1b)}$$

### Ex-1.2. Energy Values

#### 1) Ground State Energy

To determine ground state energy (or zero point energy) we can use the Equation (8) and we obtain:

$$E_0 = \frac{2\hbar^2}{mL^2} \tag{E1-2}$$

The known ground state energy is: $E_0 = \dfrac{\hbar^2 \pi^2}{2mL^2} \approx 2.465 \, \dfrac{2\hbar^2}{mL^2}$

#### 2) Excited State Energies

To determine excited state energies, we can use the equations (9):

**(a)** From (9a) we can have the symmetric state energies as:

$$E_n = \frac{2\pi^2\hbar^2}{mL^2}(n-\tfrac{1}{2})^2 = \pi^2(n-\tfrac{1}{2})^2 E_0, \qquad n = 1, 2, 3, ... \qquad (E1\text{-}3a)$$

**(b)** From (9b), we can have the antisymmetric state energies as:

$$E_n = \frac{2\pi^2\hbar^2}{mL^2}n^2 = \pi^2 n^2 E_0, \qquad n = 1, 2, 3, ... \qquad (E1\text{-}3b)$$

**(c)** From (9c) we can have the general case energies as:

$$E_n = \frac{\pi^2\hbar^2}{2mL^2}n^2 = \frac{\pi^2}{4}n^2 E_0, \qquad n = 1, 2, 3, ... \qquad (E1\text{-}3c)$$

This energy value is the same with the known energy value.

### Example 2 : Simple Harmonic Oscillator Potential

The potential energy of the simple harmonic oscillator is $U(x) = \frac{1}{2}m\omega^2 x^2$. This potential is depicted in Figure 3.

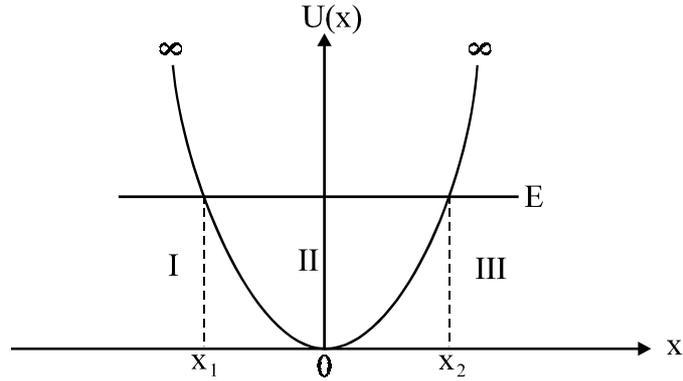

**Figure 3.** The turning points of the simple harmonic oscillator potential.

$x_1$ and $x_2$ are the roots of the equation $E = \frac{1}{2}m\omega^2 x^2$, that is $x_1 = -\sqrt{\frac{2E}{m\omega^2}}$, $x_2 = \sqrt{\frac{2E}{m\omega^2}}$.

Thus, $x_0 = 0$, $d = x_2 - x_1 = 2\sqrt{\frac{2E}{m\omega^2}}$

In the domain I and III, the potentials are infinite. Therefore, the corresponding wave functions must vanish in these domains. Namely, $\psi_1(x) = 0$ and $\psi_3(x) = 0$. In the domain II, the wave function is different from zero, namely $\psi_2(x) = \psi(x) \neq 0$. In this domain, $E > U(x)$ and $E > 0$. Thus,

$$G(x) = im_1\int\sqrt{U(x)}\,dx = im_1\int\sqrt{\tfrac{1}{2}m\omega^2 x^2}\,dx = iax^2 = iQ(x), \quad \text{with } a = \frac{m\omega}{2\hbar} \text{ and } Q(x)=ax^2$$

#### Ex-2.1. Wave Functions

From the Equations (7a) and (7b), we obtain the wave functions as:

$$\psi_n(x) = A\cos\left[\frac{(2n-1)\pi}{d}x\right]e^{-Q(x)}, \quad n = 1, 2, 3, ... \text{ (symmetric function)} \qquad (E2\text{-}1a)$$

$$\psi_n(x) = B\sin\left[\frac{2n\pi}{d}x\right]e^{-Q(x)} \qquad , \quad n = 1, 2, 3, \ldots \text{ (antisymmetric function)} \quad \text{(E2-1b)}$$

We can find the coefficients A and B by the normalisation of the wave functions. The known wave function is: $\psi_n(y) = A_n e^{-y^2/2} H_n(y)$, where $y = \sqrt{\frac{m\omega}{\hbar}}\, x$ and $H_n(y)$ Hermite Polynomials.

### Ex-2.2. Energy Values

#### 1) Ground State Energy

To determine ground state energy (or zero point energy) we can use the Equation (8) and we obtain:

$$E_0 = \frac{2\hbar^2}{m}\frac{1}{d^2}. \quad \text{Since} \quad d = 2\sqrt{\frac{2E_0}{m\omega^2}}, \quad \text{then we have} \quad E_0 = \frac{1}{2}\hbar\omega \qquad \text{(E2-2)}$$

#### 2) Excited State Energies

To determine excited state energies, we can use the equations (9):

**(a)** From (9a), we can have the symmetric state energies as follows:

$$E_n = \frac{2\hbar^2\pi^2}{md^2}\left(n-\tfrac{1}{2}\right)^2 = \frac{2\hbar^2\pi^2}{m4\dfrac{2E_n}{m\omega^2}}\left(n-\tfrac{1}{2}\right)^2. \quad \text{From this equation we get:}$$

$$E_n = \frac{\pi}{2}(n-\tfrac{1}{2})\hbar\omega = \pi(n-\tfrac{1}{2})E_0, \qquad n = 1, 2, 3, \ldots \qquad \text{(E2-3a)}$$

**(b)** From (9b), we can have the antisymmetric state energies as follows:

$$E_n = \frac{2\hbar^2\pi^2}{md^2}n^2 = \frac{2\hbar^2\pi^2}{m4\dfrac{2E_n}{m\omega^2}}n^2. \quad \text{From this equation we have:}$$

$$E_n = \frac{\pi}{2}n\hbar\omega = \pi n E_0, \qquad n = 1, 2, 3, \ldots \qquad \text{(E2-3b)}$$

**(c)** From (9c), we can have the general case energies as follows:

$$E_n = \frac{\hbar^2\pi^2}{2md^2}n^2 = \frac{\hbar^2\pi^2}{2m4\dfrac{2E_n}{m\omega^2}}n^2. \quad \text{From this equation we have:} \qquad E_n^2 =$$

$$E_n = \frac{\pi}{4}n\hbar\omega = \frac{\pi}{2}nE_0, \qquad \text{with} \quad n = 1, 2, 3, \ldots \qquad \text{(E2-3c)}$$

The known energy value is: $E_n = (n+\frac{1}{2})\hbar\omega = 2(n+\frac{1}{2})E_0$, with $n = 0, 1, 2, \ldots$

### Example 3. Infinitely High Trigonometric Potential Well

We consider the potential energy of the particle, U(x), so that

$\cot g^2\left(\dfrac{\pi x}{a}\right)$, with $U_0 > 0$, $a > 0$ and $0 < x < a$.

This potential is depicted in Figure 4.

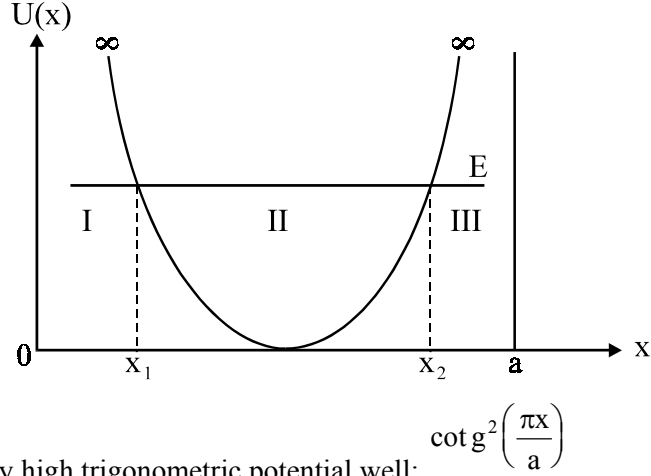

**Figure 4.** An infinitely high trigonometric potential well: $\cot g^2\left(\dfrac{\pi x}{a}\right)$.

$x_1$ and $x_2$ are the positive roots of the equation $E = U(x) = U_0 \cot g^2\left(\dfrac{\pi x}{a}\right)$. Thus

$$x_1 = \dfrac{a}{\pi}[\pi - arc \cot g(\sqrt{\dfrac{E}{U_0}})],\ x_2 = \dfrac{a}{\pi} arc \cot g(\sqrt{\dfrac{E}{U_0}}),\ x_0 = \dfrac{x_1 + x_2}{2} = \dfrac{a}{2},$$

$$d = x_2 - x_1 = -a[1 - \dfrac{2}{\pi} arc \cot(\sqrt{\dfrac{E}{U_0}})]$$

In the domain I and III, the potentials are infinite. Therefore, the corresponding wave functions must vanish in these domains. Namely, $\psi_1(x) = 0$ and $\psi_3(x) = 0$. In the domain II, the wave function is different from zero, that is, $\psi_2(x) = \psi(x) \neq 0$. In this domain, $E > U(x)$ and $E > 0$. Thus,

$$G(x) = i m_1 \int \sqrt{U(x)} dx = i m_1 \sqrt{U_0} \dfrac{a}{\pi} \ln\left[\sin(\dfrac{\pi x}{a})\right] = i Q(x),$$

with $\quad Q(x) = \sqrt{\dfrac{2m}{\hbar^2} U_0}\ \dfrac{a}{\pi} \ln\left[\sin\left(\dfrac{\pi x}{a}\right)\right]$

**Ex-3.1. Wave Functions**

From the Equations (7a) and (7b), we obtain the wave functions as follows:

$$\psi_n(x) = A\cos\left[\dfrac{(2n-1)\pi}{d}(x - \dfrac{a}{2})\right]e^{-Q(x)}, \quad n = 1, 2, 3, \ldots \text{ (symmetric function)} \quad \text{(E3-1a)}$$

$$\psi_n(x) = B \sin\left[\frac{2n\pi}{d}(x - \frac{a}{2})\right] e^{-Q(x)} \quad , \quad n = 1, 2, 3, \ldots \text{ (antisymmetric function)} \quad \text{(E3-1b)}$$

We can find the coefficients A and B by the normalisation of the wave functions.

**Ex-3.2. Energy Values**

**1) Ground State Energy**

To find ground state energy (or zero point energy) we can use the equation (8) and we got:

$$E_0 = \frac{2\hbar^2}{ma^2} \frac{1}{\left[1 - \frac{2}{\pi} \text{arc cot } g\left(\sqrt{\frac{E_0}{U_0}}\right)\right]^2} \quad \text{(E3-2)}$$

**2) Excited State Energies**

To determine excited state energies, we can use the equations (9).

**(a)** From (9a) we can have the symmetric state energies as:

$$E_n = \frac{2\hbar^2 \pi^2}{m d_n^2} (n - \tfrac{1}{2})^2 \quad , \quad n = 1, 2, 3, \ldots \quad \text{(E3-3a)}$$

**(b)** From (9b) we can have the antisymmetric state energies as:

$$E_n = \frac{2\hbar^2 \pi^2}{m d_n^2} n^2 \quad , \quad n = 1, 2, 3, \ldots \quad \text{(E3-3b)}$$

**(c)** From (9c) we can have the general case energies as:

$$E_n = \frac{\hbar^2 \pi^2}{2 m d_n^2} n^2 \quad , \quad n = 1, 2, 3, \ldots \quad \text{(E3-3c)}$$

Where $d_n^2 = a^2 \left[1 - \frac{2}{\pi} \text{arc cot } g\left(\sqrt{\frac{E_n}{U_0}}\right)\right]^2$.

We should solve these equations by numerical methods.

**Example 4 : Infinitely High V-Form Potential Well**

We consider the potential energy of the particle, $U(x) = U_0 |x|$, with $U_0 > 0$ constant. This potential is depicted in Figure 5.

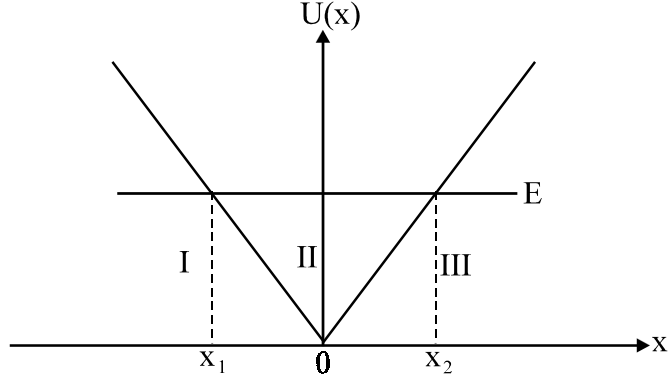

**Figure 5.** An infinitely high trigonometric potential well: $U(x) = U_0 |x|$, with $U_0 > 0$ constant.

This potential function can be written as

$U(x) = -U_0 x$      for     $x < 0$,

$U(x) = U_0 x$      for     $x > 0$.

$x_1$ and $x_2$ are the roots of the equation $U(x) = U_0 |x| = E$. Thus,

$$x_1 = -\frac{E}{U_0} \quad \text{and} \quad x_2 = \frac{E}{U_0}, \quad x_0 = \frac{x_1 + x_2}{2} = 0, \quad d = x_2 - x_1 = 2\frac{E}{U_0}$$

In the domains I and III, the potentials are infinite. Therefore, the corresponding wave functions must vanish in these domains. Namely, $\psi_1(x) = 0$ and $\psi_3(x) = 0$. In the domain II, the wave function is different from zero, namely $\psi_2(x) = \psi(x) \neq 0$. In this domain, $E > U(x)$ and $E > 0$. Thus,

$$G(x) = i m_1 \int \sqrt{U(x)} dx = i Q(x), \quad \text{with} \quad Q(x) = m_1 \sqrt{U_0} \int \sqrt{|x|} \, dx = \sqrt{\frac{2m}{\hbar^2} U_0} \, \frac{2}{3} x^{3/2} \quad (x > 0)$$

**Ex-4.1. Wave Functions**

From the Equations (7a) and (7b), we obtain the wave functions as:

$$\psi_n(x) = A \cos\left[\frac{(2n-1)\pi}{d} x\right] e^{-Q(x)}, \quad n = 1, 2, 3, \ldots \text{(symmetric function)} \quad \text{(E4-1a)}$$

$$\psi_n(x) = B \sin\left[\frac{2n\pi}{d} x\right] e^{-Q(x)}, \quad n = 1, 2, 3, \ldots \text{(antisymmetric function)} \quad \text{(E4-1b)}$$

We can find the coefficients A and B by the normalisation of the wave functions.

**Ex-4.2. Energy Values**

**1) Ground State Energy**

To determine ground state energy (or zero point energy) we can use the Equation (8) and we obtain:

$$E_0 = \frac{2\hbar^2}{m} \frac{1}{d^2} \; . \; \text{Since} \; d = \frac{2E_0}{U_0}, \; \text{then we have} \; E_0 = \frac{2\hbar^2}{m} \frac{U_0^2}{4E_0^2} \to E_0^3 = \frac{\hbar^2 U_0^2}{2m}$$

Or from this equation we have got:

$$E_0 = \left[\frac{\hbar^2 U_0^2}{2m}\right]^{\frac{1}{3}} = \left(\frac{1}{2}\right)^{\frac{1}{3}}\left[\frac{\hbar^2 U_0^2}{m}\right]^{\frac{1}{3}} \approx 0.794 \left[\frac{\hbar^2 U_0^2}{m}\right]^{\frac{1}{3}} \tag{E4-2}$$

The known ground state energy (with variational method) is:

$$E_0 = \frac{3}{2}\left(\frac{1}{2\pi}\right)^{\frac{1}{3}}\left[\frac{\hbar^2 U_0^2}{m}\right]^{\frac{1}{3}} \approx 0.813 \left[\frac{\hbar^2 U_0^2}{m}\right]^{\frac{1}{3}}$$

**2) Excited State Energies**

To determine excited state energies, we can use the equations (9)

**(a)** From (9a) we can have the symmetric state energies as:

$$E_n^3 = \frac{\hbar^2 \pi^2}{2m} U_0^2 (n-\frac{1}{2})^2 \quad \text{or} \quad E_n = \left[\frac{\hbar^2 \pi^2}{2m} U_0^2 (n-\frac{1}{2})^2\right]^{1/3}$$

$$E_n = \left[\pi^2 (n-\frac{1}{2})^2\right]^{1/3} E_0, \qquad n = 1, 2, 3, \ldots \tag{E4-3a}$$

**(b)** From (9b), we can have the antisymmetric state energies as:

$$E_n^3 = \frac{\hbar^2 \pi^2}{2m} U_0^2 n^2 \quad \text{or} \quad E_n = \left[\frac{\hbar^2 \pi^2}{2m} U_0^2 n^2\right]^{1/3} = \left[\pi^2 n^2\right]^{1/3}\left[\frac{\hbar^2 U_0^2}{2m}\right]^{1/3}$$

$$E_n = \left[\pi^2 n^2\right]^{1/3} E_0, \qquad n = 1, 2, 3, \ldots \tag{E4-3b}$$

**(c)** From (9c) we can have the general case energies as:

$$E_n^3 = \frac{\hbar^2 \pi^2}{8m} U_0^2 n^2 = \frac{\pi^2}{4} n^2 \frac{\hbar^2 U_0^2}{2m}, \quad \text{or} \quad E_n = \left[\frac{\pi^2}{4} n^2\right]^{1/3}\left[\frac{\hbar^2 U_0^2}{2m}\right]^{1/3}$$

$$E_n = \left[\frac{\pi^2}{4} n^2\right]^{1/3} E_0, \qquad n = 1, 2, 3, \ldots \tag{E4-3c}$$

**Example 5. Infinitely High Parabolic Potential Well**

We consider the potential energy of the particle $U(x) = U_0 \left(\frac{a}{x} - \frac{x}{a}\right)^2$, where $U_0 > 0$, $a > 0$, $x > 0$. This potential is depicted in Figure 6.

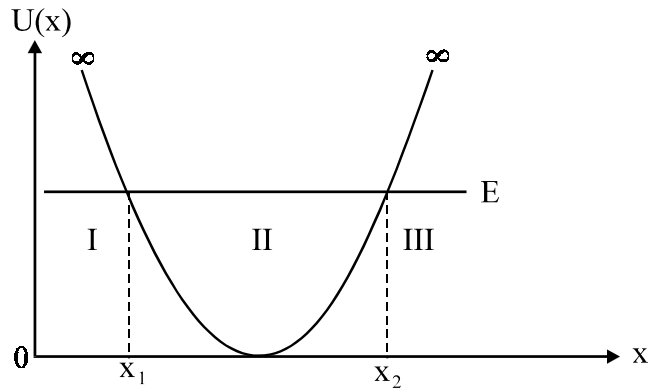

**Figure 6.** An infinitely high parabolic potential well: $U(x) = U_0 \left( \dfrac{a}{x} - \dfrac{x}{a} \right)^2$.

$x_1$ and $x_2$ are the positive roots of the equation $E = U_0 \left( \dfrac{a}{x} - \dfrac{x}{a} \right)^2$. Thus,

$$x_1 = \frac{a}{2}\left[ -\sqrt{\frac{E}{U_0}} + \sqrt{\frac{E}{U_0} + 4} \right], \quad x_2 = \frac{a}{2}\left[ \sqrt{\frac{E}{U_0}} + \sqrt{\frac{E}{U_0} + 4} \right], \quad d = x_2 - x_1 = a\sqrt{\frac{E}{U_0}},$$

$$d^2 = a^2 \frac{E}{U_0}, \quad x_0 = \frac{x_1 + x_2}{2} = \frac{a}{2}\sqrt{\frac{E}{U_0} + 4}, \quad x_1 = x_0 - d/2, \quad x_2 = x_0 + d/2,$$

In the domains I and III, the potentials are infinite. Therefore, the corresponding wave functions must vanish in these domains. Namely, $\psi_1(x) = 0$ and $\psi_3(x) = 0$. In the domain II, the wave function is different from zero, namely $\psi_2(x) = \psi(x) \neq 0$. In this domain, $E > U(x)$ and $E > 0$. Thus,

$$G(x) = i m_1 \int \sqrt{U(x)} dx = i m_1 \int \sqrt{U_0 \left( \frac{a}{x} - \frac{x}{a} \right)^2} dx = i m_1 \sqrt{U_0} \left[ \int \frac{a}{x} dx - \int \frac{x}{a} dx \right]$$

$$G(x) = i m_1 \sqrt{U_0} \left[ a\ln(x) - \frac{x^2}{2a} \right] = i Q(x), \quad \text{where} \quad Q(x) = \sqrt{\frac{2m}{\hbar^2} U_0} \left[ a\ln(x) - \frac{x^2}{2a} \right]$$

**Ex-5.1. Wave Functions**

From the Equations (7a) and (7b), we have the wave functions as:

$$\psi_n(x) = A\cos\left[ \frac{(2n-1)\pi}{d}(x - x_0) \right] e^{-Q(x)}, \quad n = 1, 2, 3, \ldots \text{ (symmetric function) (E5-1a)}$$

$$\psi_n(x) = B\sin\left[ \frac{2n\pi}{d}(x - x_0) \right] e^{-Q(x)}, \quad n = 1, 2, 3, \ldots \text{ (antisymmetric function) (E5-1b)}$$

We can find the coefficients A and B by the normalisation of the wave functions.

**Ex-5.2. Energy Values**

**1) Ground State Energy**

To determine ground state energy (or zero point energy) we can use the Equation (8) and we obtain:

$$E = \frac{2\hbar^2}{m}\frac{1}{d^2}. \quad \text{Since} \quad d^2 = a^2 \frac{E}{U_0}, \quad \text{we have} \quad E_0 = \frac{2\hbar^2}{m}\frac{U_0}{a^2 E_0} \rightarrow E_0^2 = \frac{2\hbar^2}{ma^2} U_0$$

$$E_0 = \sqrt{\frac{2\hbar^2}{ma^2} U_0} \tag{E5-2}$$

**2) Excited State Energies**

To determine excited state energies, we can use the equations (9).

**(a)** From (9a), we can have the symmetric state energies as:

$$E_n^2 = \frac{2\pi^2\hbar^2 U_0}{ma^2}(n-\tfrac{1}{2})^2 \quad \text{or} \quad E_n = \pi\sqrt{\frac{2\hbar^2}{ma^2}U_0}(n-\tfrac{1}{2})$$

$$E_n = \pi(n-\tfrac{1}{2})E_0 \,, \qquad n = 1, 2, 3, ... \qquad (E5\text{-}3a)$$

**(b)** From (9b), we can have the antisymmetric state energies as:

$$E_n^2 = \frac{2\pi^2\hbar^2 U_0}{ma^2}n^2 \quad \text{or} \quad E_n = \pi\sqrt{\frac{2\hbar^2}{ma^2}U_0}\, n$$

$$E_n = \pi n E_0 \,, \qquad n = 1, 2, 3, ... \qquad (E5\text{-}3b)$$

**(c)** From (9c), we can have the general case energies as:

$$E_n^2 = \frac{\pi^2 2\hbar^2 U_0}{4ma^2}n^2 \quad \text{or} \quad E_n = \frac{\pi}{2}\sqrt{\frac{2\hbar^2}{ma^2}U_0}\, n$$

$$E_n = \frac{\pi}{2} n E_0 \,, \qquad n = 1, 2, 3, ... \qquad (E5\text{-}3c)$$

**Example 6. Potential $U(x) = a x^2 + b / x^2$**

We consider the potential energy of the particle as $U(x) = ax^2 + \dfrac{b}{x^2}$, $a$ and $b$ are positive constants. This potential is depicted in Figure 7.

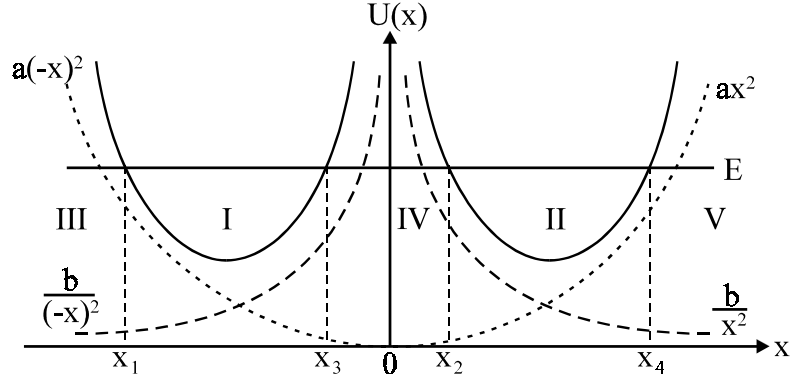

**Figure 7.** Turning points of the potential $U(x) = ax^2 + \dfrac{b}{x^2}$.

The roots of the equation $E = ax^2 + \dfrac{b}{x^2}$ are:

$$x_3 = -\sqrt{\frac{E-\sqrt{\Delta}}{2a}}, \quad x_2 = \sqrt{\frac{E-\sqrt{\Delta}}{2a}}, \quad x_1 = -\sqrt{\frac{E+\sqrt{\Delta}}{2a}}, \quad x_4 = \sqrt{\frac{E+\sqrt{\Delta}}{2a}}$$

where $\Delta = E^2 - 4ab$

$$d_1 = x_3 - x_1 = -\sqrt{\frac{E+\sqrt{\Delta}}{2a}} + \sqrt{\frac{E-\sqrt{\Delta}}{2a}}, \quad d_1^2 = (x_3-x_1)^2 = \frac{E}{a} - \frac{1}{a}\sqrt{E^2+\Delta} = \frac{E}{a} - 2\sqrt{\frac{b}{a}}$$

$$d_2 = x_4 - x_2 = \sqrt{\frac{E+\sqrt{\Delta}}{2a}} - \sqrt{\frac{E-\sqrt{\Delta}}{2a}}, \quad d_2^2 = (x_4-x_2)^2 = \frac{E}{a} - \frac{1}{a}\sqrt{E^2+\Delta} = \frac{E}{a} - 2\sqrt{\frac{b}{a}}$$

$$x_{01} = \frac{x_1 + x_3}{2} = -\frac{1}{2}\left[\sqrt{\frac{E+\sqrt{\Delta}}{2a}} + \sqrt{\frac{E-\sqrt{\Delta}}{2a}}\right], \quad x_{02} = \frac{x_2 + x_4}{2} = \frac{1}{2}\left[\sqrt{\frac{E+\sqrt{\Delta}}{2a}} + \sqrt{\frac{E-\sqrt{\Delta}}{2a}}\right]$$

From these equations, we can easily obtain :

$$d_1 = -d_2 = d, \quad x_{01} = -x_{02} = x_0, \quad d_1^2 = d_2^2 = d^2 = \frac{E}{a} - 2\sqrt{\frac{b}{a}}$$

In the domain III, IV and V, the potentials are infinite. Therefore, the corresponding wave functions must vanish in these domains. Namely, $\psi_3(x) = 0$, $\psi_4(x) = 0$ and $\psi_5(x) = 0$. In the domain I and II, the total energy is greater than the potential energy. Thus,

$$G(x) = i m_1 \int \sqrt{U(x)}\, dx = i m_1 \int \sqrt{ax^2 + \frac{b}{x^2}}\, dx = i Q(x)$$

with $\quad Q(x) = Q(x) = \frac{m_1}{2}\left\{\sqrt{ax^4 + b} - \sqrt{b}\ln\left[\frac{\sqrt{b} + \sqrt{ax^4 + b}}{\sqrt{a}x^2}\right]\right\}$

Since the potential energy is symmetric in the domains I and II, the corresponding wave functions are the same, namely, $\psi_1(x) = \psi_2(x) = \psi(x)$. This function must vanish at $x = \pm\infty$ and $x = 0$.

**Ex-6.1. Wave Functions**

From the Equations (7a) and (7b), we obtain the wave functions as:

$$\psi_n(x) = A\cos\left[\frac{(2n-1)\pi}{d}(x - x_0)\right]e^{-Q(x)}, \quad n = 1, 2, 3, \ldots \text{ (symmetric function)} \quad (E6\text{-}1a)$$

$$\psi_n(x) = A\sin\left[\frac{2\pi n}{d}(x - x_0)\right]e^{-Q(x)}, \quad n = 1, 2, 3, \ldots \text{(antisymmetric function)} \ (E6\text{-}1b)$$

We can find the coefficients A and B by the normalisation of the wave functions.

**Ex-6.2. Energy Values**

**1) Ground State Energy**

To determine ground state energy (or zero point energy) we can use the Equation (8) and we obtain:

$$E = \frac{2\hbar^2}{m}\frac{1}{d^2} = \frac{2\hbar^2}{m}\frac{1}{\frac{E}{a} - 2\sqrt{\frac{b}{a}}} \quad \text{or} \quad E^2 - 2\sqrt{ab}E - \frac{2\hbar^2 a}{m} = 0. \quad \text{From this equation}$$

$$E_{1,2} = \sqrt{ab} \mp \sqrt{ab + \frac{2a\hbar^2}{m}}$$

Since $E > 0$ and the second term is greater than the first one, we should take only the (+) sign. Therefore, ground state energy is

$$E_0 = \sqrt{ab} + \sqrt{ab + \frac{2a\hbar^2}{m}} \qquad (E6\text{-}2)$$

**2) Excited State Energies**

To determine excited state energies, we can use the equations (9).

**(a)** From (9a), we can have the symmetric state energies as:

$$\frac{2m}{\hbar^2} E \left[ \frac{E}{a} - 2\sqrt{\frac{b}{a}} \right] = 4\pi^2 \left(n - \tfrac{1}{2}\right)^2 \quad \text{or} \quad E^2 - 2\sqrt{ab}\,E - \frac{2\pi^2 \hbar^2 a}{m}\left(n - \tfrac{1}{2}\right)^2 = 0$$

From this equation, we get $E_{1,2} = \sqrt{ab} \mp \sqrt{ab + \frac{2\pi^2 \hbar^2 a}{m}\left(n - \tfrac{1}{2}\right)^2}$. Since E > 0 and the second term is greater than the first one, we should take only the (+) sign. Therefore, excited symmetric state energy is

$$E_n = \sqrt{ab} + \sqrt{ab + \frac{2\pi^2 \hbar^2 a}{m}\left(n - \tfrac{1}{2}\right)^2}, \qquad n = 1, 2, 3, \ldots \qquad (E6\text{-}3a)$$

**(b)** From (9b), we can have the antisymmetric state energies as:

$$\frac{2m}{\hbar^2} E \left[ \frac{E}{a} - 2\sqrt{\frac{b}{a}} \right] = 4\pi^2 n^2 \quad \text{or} \quad E^2 - 2\sqrt{ab}\,E - \frac{2\pi^2 \hbar^2 a}{m} n^2 = 0$$

From this equation, we get $E_{1,2} = \sqrt{ab} \mp \sqrt{ab + \frac{2\pi^2 \hbar^2 a}{m} n^2}$. Since E > 0 and the second term is greater than the first one, we should take only the (+) sign. Therefore, the excited antisymmetric state energy is

$$E_n = \sqrt{ab} + \sqrt{ab + \frac{2\pi^2 \hbar^2 a}{m} n^2}, \qquad n = 1, 2, 3, \ldots \qquad (E6\text{-}3b)$$

**(c)** From (9c), we can have the general case energies as:

$$\frac{2m}{\hbar^2} E \left[ \frac{E}{a} - 2\sqrt{\frac{b}{a}} \right] = \pi^2 n^2 \quad \text{or} \quad E^2 - 2\sqrt{ab}\,E - \frac{\pi^2 \hbar^2 a}{2m} n^2 = 0$$

From this equation, we get $E_{1,2} = \sqrt{ab} \mp \sqrt{ab + \frac{\pi^2 \hbar^2 a}{2m} n^2}$. Since E > 0 and the second term is greater than the first one, we should take only the (+) sign. Therefore, the general case energy is

$$E_n = \sqrt{ab} + \sqrt{ab + \frac{\pi^2 \hbar^2 a}{2m} n^2}, \qquad n = 1, 2, 3, \ldots \qquad (E6\text{-}3c)$$

## 4. CONCLUSION

By using our method, we solved the Schrödinger equation for a particle in an infinitely high potential well of arbitrary form and we found two different solutions. One of them is symmetric function and the other is antisymmetric function. They are given by the equations (7a) and (7b), respectively. Their corresponding energy values are also given by (9a) and (9b), respectively. Ground state energy is given by the equation (8). From these expressions, we observe that these functions are periodic and they are similar to each other for all infinitely high potential wells.

# PART III. SOLUTION OF THE SCHRÖDINGER EQUATION IN ONE DIMENSION BY A SIMPLE METHOD FOR A SIMPLE STEP POTENTIAL

H. H. Erbil [a]

Ege University, Science Faculty, Physics Department    Bornova - IZMIR 35100, TURKEY

The coefficients of the transmission and reflection for the simple-step barrier potential were calculated by a simple method. Their values were entirely different from those often encountered in the literature. Especially in the case that the total energy is equal to the barrier potential, the value of 0.20 for the reflection coefficient was obtained whereas this is one in the literature. This may be considered as an interesting point.

PACS: 03.65.-w and 03.65.Ge

Keywords: simple step potential

## 1. INTRODUCTION

The time-independent Schrödinger equation in one dimension is given as follows:

$$\frac{d^2\psi(x)}{dx^2} + \frac{2m}{\hbar^2}[E - U(x)]\psi(x) = 0$$

Where, E and U(x) are total and potential energy, respectively. In the previous study [1], we have generally solved this Schrödinger equation in one dimension and we obtained some functions and one of them is as below:

$$\psi(x) = A e^{kx + iG(x)} + B e^{-kx - iG^*(x)}$$

Where, (a)   For E > U(x),   $k = i m_1 \sqrt{E} = iK$,     $G(x) = i m_1 \int \sqrt{U(x)}\, dx$

(b)   For E < U(x),   $k = m_1 \sqrt{E} = K$,     $G(x) = m_1 \int \sqrt{U(x)}\, dx$

with $m_1 = \sqrt{\frac{2m}{\hbar^2}}$. A and B are integral constants to be determined by the boundary conditions.

Let us apply this function to the simple step barrier potential

## 2. WAVE FUNCTIONS, COEFFICIENTS OF TRANSMISSION AND REFLECTION

Let us consider the simple step potential function shown in Figure 1.

The potential function is zero for x < 0 and is constant $U_0$ for x ≥ 0. Namely,

$U(x) = 0$    for    $x < 0$

$U(x) = +U_0$    for    $x \geq 0$

---

[a]  E-mail: erbil@sci.ege.edu.tr    Phone: +90 232 388 4000 / 2379    Fax: +90 232 388 1036

26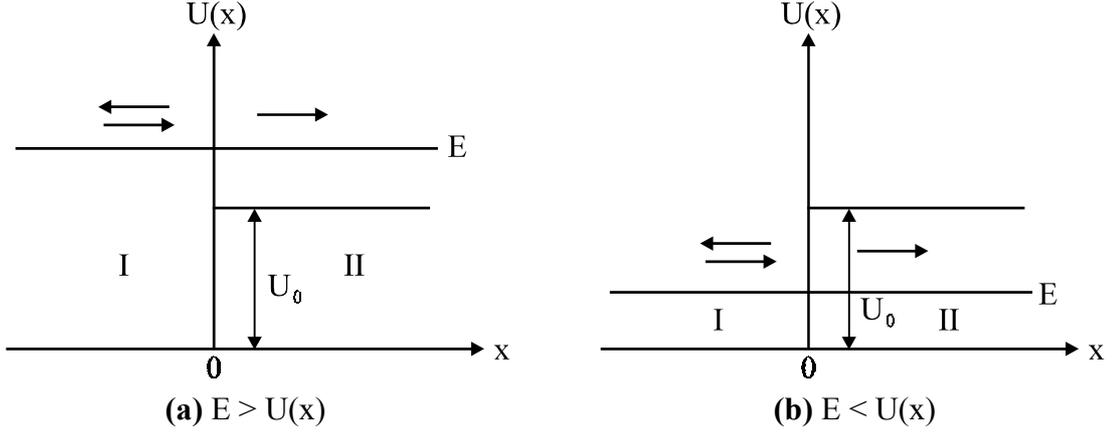

**(a)** E > U(x)     **(b)** E < U(x)

**Figure 1.** The simple step potential (a) The case E > U(x)  (b) The case E < U(x).

The incident beam comes from x = − ∞. To construct $\psi(x)$, we divide x axis into two domains: region I and region II, depicted in Figure 1. We consider two cases:

**1) The case E > U(x)**

In this case, in region I, U(x) = 0,   E > U(x),   k = i K,   G(x) = 0

$$\psi_1(x) = A_1 e^{iKx} + B_1 e^{-iKx} \quad \text{with} \quad K = \sqrt{\frac{2m}{\hbar^2}E} \tag{1a}$$

In region II,   U(x) = $U_0$,   E > U(x)

$$k = iK \quad \text{and} \quad G(x) = i m_1 \int \sqrt{U(x)}\, dx = i m_1 \int \sqrt{U_0}\, dx = i a x = i Q(x)$$

with   $a = \sqrt{\frac{2m}{\hbar^2}U_0}$   and   Q(x) = a x,   Q(x) is real.

$$\psi_2(x) = A_2 e^{iKx+iG(x)} + B_2 e^{-iKx-iG^*(x)}, \text{ thus,}$$

$$\psi_2(x) = A_2 e^{iKx-Q(x)} + B_2 e^{-iKx-Q(x)} = \left[A_2 e^{iKx} + B_2 e^{-iKx}\right] e^{-Q(x)} \tag{1b}$$

Since the term $B_2 e^{-iKx}$ represents a wave emanating from the right (x = + ∞ in Figure 1) and there is no such a wave, we may conclude that $B_2 = 0$. The interpretation of the remaining $A_1$, $B_1$ and $A_2$ terms is as follows: $A_1 e^{iKx}$ represents the incident wave; $B_1 e^{-iKx}$ represents the reflected wave; and $A_2 e^{iKx-Q(x)}$ represents the transmitted wave. Since any wave function and its first derivative are continuous at the point x = 0, where $\psi_1(x)$ and $\psi_2(x)$ join, it is required that

$$\psi_1(0) = \psi_2(0), \quad \psi_1'(0) = \psi_2'(0)$$

These equalities give the relations below:

$$A_1 + B_1 = A_2, \quad iK[A_1 - B_1] = [iK - a] A_2$$

If we solve these equations, we obtain:

$$B_1 = \frac{a}{[2iK - a]} A_1 \quad \text{and} \quad A_2 = \frac{2iK}{[2iK - a]} A_1 \tag{2}$$

The transmission coefficient T and the reflection coefficient R are defined as

$$T = \frac{A_2^* e^{-iKx-Q(x)} \cdot A_2 e^{iKx-Q(x)}}{A_1^* e^{-iKx} \cdot A_1 e^{iKx}} \quad \text{and} \quad R = \frac{B_1^* e^{iKx} \cdot B_1 e^{-iKx}}{A_1^* e^{-iKx} \cdot A_1 e^{iKx}} \quad (3)$$

From (3), we obtain :

$$T = \frac{A_2^* \cdot A_2 e^{-2Q(x)}}{A_1^* \cdot A_1} \quad \text{and} \quad R = \frac{B_1^* \cdot B_1}{A_1^* \cdot A_1} \quad (4)$$

Substituting the values of $B_1$ and $A_2$ into (4) yields

$$T = \frac{4K^2}{4K^2 + a^2} e^{-2Q(x)} = \frac{4E}{4E + U_0} e^{-2ax} = T_0 \, e^{-2ax} \quad \text{with} \quad T_0 = \frac{4E}{4E + U_0} \quad (5a)$$

$$R = \frac{a^2}{4K^2 + a^2} = \frac{U_0}{4E + U_0} \quad (5b)$$

The graphics of T and R are shown in Figure 2.

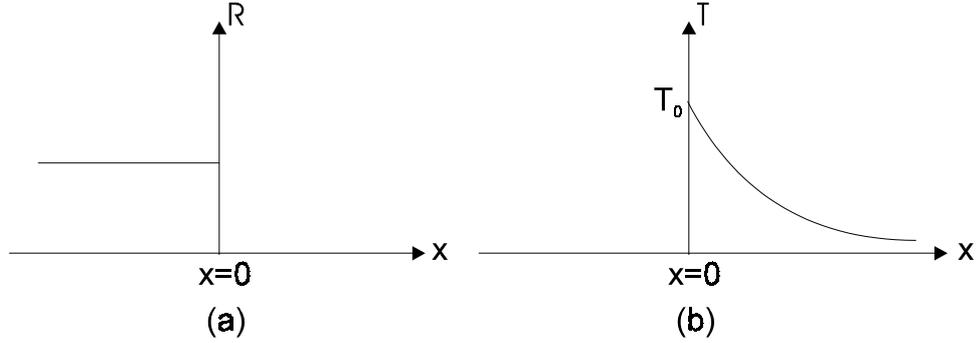

(a)　　　　　　　　　　　　(b)

**Figure 2.** The coefficients of reflection and transmission **(a)** R is the reflection coefficient **(b)** T is the transmission coefficient, where $T_0 = 1 - R = \frac{4E}{4E + U_0}$, $R = \frac{U_0}{4E + U_0}$.

From the equations (5), we see that

1) For $x = 0$, $T + R = 1$
2) For $x = \infty$, $T = 0$
3) For $E = U_0$, $T = \frac{4}{5} e^{-2ax} = 0.80 \, e^{-2ax}$, $R = \frac{1}{5} = 0.20$
4) For $E \gg U_0$, $T \approx 1 \cdot e^{-2ax}$ and $R \approx 0$

**2) The case $E < U(x)$**

In this case, in region I, $U(x) = 0$, $E > U(x)$, $k = iK$, $G(x) = 0$

$$\psi_1(x) = A_1 e^{iKx} + B_1 e^{-iKx} \quad \text{with} \quad K = \sqrt{\frac{2m}{\hbar^2} E} \quad (6a)$$

In region II, $U(x) = U_0$, $E < U(x)$

$$k = K \quad \text{and} \quad G(x) = m_1 \int \sqrt{U(x)}\, dx = m_1 \int \sqrt{U_0}\, dx = a\, x = Q(x)$$

with $\sqrt{\dfrac{2m}{\hbar^2} U_0}$ and $Q(x) = a\, x$

$$\psi_2(x) = A_2 e^{Kx + iQ(x)} + B_2 e^{-Kx - iQ(x)} \tag{6b}$$

Since the term $B_2 e^{-Kx - iQ(x)}$ represents a wave emanating from the right (x = +∞ in Figure 1) and there is no a such wave, we may conclude that $B_2 = 0$. Since any wave function and its first derivative are continuous at the point x = 0, where $\psi_1(x)$ and $\psi_2(x)$ join, it is required that

$$\psi_1(0) = \psi_2(0), \quad \psi_1'(0) = \psi_2'(0)$$

These equalities give the relations below:

$$A_1 + B_1 = A_2 \quad \text{and} \quad i K [A_1 - B_1] = [K + i a] A_2 \tag{7}$$

If we solve these equations, we obtain :

$$B_1 = \frac{[i(K - a) - K]}{[i(K + a) + K]} A_1 \quad \text{and} \quad A_2 = \frac{2iK}{K + i(K - a)} A_1 \tag{8}$$

Transmission coefficient T and the reflection coefficient R are:

$$T = \frac{A_2^* e^{Kx - iQ(x)} \cdot A_2 e^{Kx + iQ(x)}}{A_1^* e^{-iKx} \cdot A_1 e^{iKx}} = \frac{4K^2}{K^2 + (K + a)^2} e^{2Kx} = \frac{4E}{E + \left[\sqrt{E} + \sqrt{U_0}\right]^2} e^{2Kx} \tag{9a}$$

$$R = \frac{B_1^* e^{iKx} \cdot B_1 e^{-iKx}}{A_1^* e^{-iKx} \cdot A_1 e^{iKx}} = \frac{K^2 + (K - a)^2}{K^2 + (K + a)^2} = \frac{E + \left[\sqrt{E} - \sqrt{U_0}\right]^2}{E + \left[\sqrt{E} + \sqrt{U_0}\right]^2} \tag{9b}$$

From the equality (9a), we see that T = ∞ for x = ∞. Since T must be equal to or less than 1, namely T ≤ 1, the value of T in (9a) is not possible. Thus, $A_2$ must vanish. In any case, in order that the function given by (6b), $\psi_2(x)$, must vanish at x = +∞, $A_2$ should be zero. If we put $A_2 = 0$, into (7), we get $B_1 = -A_1$ or $B_1 = A_1$. Using these values, we obtain :

$$R = 1 \quad \text{and} \quad T = 0 \tag{10}$$

There is total a reflection, hence the transmission must be zero. From (9b), we can see that the value of R is 0.20 for $E = U_0$ and this value is the same as in the both cases of $E > U_0$ and $E < U_0$.

### 3. CONCLUSION

Using our method, we calculated the coefficients of the transmission and reflection for simple step barrier potential and found values different from those encountered in literature. In particularly, in the case $E = U_0$, we have got the value of 0.20 for the reflection coefficient, whereas we find that this value is one in the literature. This may be appreciated as an interesting point.